\documentclass[a4paper,11pt]{article}
\usepackage{jheppub, bm, color} 
\usepackage{amssymb,amsfonts,slashed,amsthm,amsmath,graphicx, soul, empheq}
\usepackage[caption=false]{subfig}
\usepackage{float}
\usepackage{subcaption}  
\usepackage{placeins}
\usepackage{mathrsfs}
\usepackage{booktabs}

\begin{document}

\newcommand{\vev}[1]{ \left\langle {#1} \right\rangle }
\newcommand{\bra}[1]{ \langle {#1} | }
\newcommand{\ket}[1]{ | {#1} \rangle }
\newcommand{\eV}{ \ {\rm eV} }
\newcommand{\KeV}{ \ {\rm keV} }
\newcommand{\MeV}{\  {\rm MeV} }
\newcommand{\GeV}{\  {\rm GeV} }
\newcommand{\TeV}{\  {\rm TeV} }
\newcommand{\1}{\mbox{1}\hspace{-0.25em}\mbox{l}}
\newcommand{\Red}[1]{{\color{red} {#1}}}

\newcommand{\lmk}{\left(}  
\newcommand{\rmk}{\right)}
\newcommand{\lkk}{\left[}  
\newcommand{\rkk}{\right]}
\newcommand{\lhk}{\left \{ }  
\newcommand{\rhk}{\right \} }
\newcommand{\del}{\partial}  
\newcommand{\la}{\left\langle} 
\newcommand{\ra}{\right\rangle}
\newcommand{\half}{\frac{1}{2}}

\newcommand{\bea}{\begin{array}}
\newcommand{\eea}{\end{array}}
\newcommand{\beq}{\begin{eqnarray}}
\newcommand{\eeq}{\end{eqnarray}}
\newcommand{\eq}[1]{Eq.~(\ref{#1})}

\newcommand{\dd}{\mathrm{d}}
\newcommand{\Mpl}{M_{\rm Pl}}
\newcommand{\mg}{m_{3/2}}
\newcommand{\abs}[1]{\left\vert {#1} \right\vert}
\newcommand{\mphi}{m_{\phi}}
\newcommand{\Hz}{\ {\rm Hz}}
\newcommand{\for}{\quad \text{for }}
\newcommand{\Min}{\text{Min}}
\newcommand{\Max}{\text{Max}}
\newcommand{\Kahler}{K\"{a}hler }
\newcommand{\cphi}{\varphi}
\newcommand{\Tr}{\text{Tr}}
\newcommand{\diag}{{\rm diag}}

\newcommand{\SUf}{SU(3)_{\rm f}}
\newcommand{\Upq}{U(1)_{\rm PQ}}
\newcommand{\Zpq}{Z^{\rm PQ}_3}
\newcommand{\Cpq}{C_{\rm PQ}}
\newcommand{\ubar}{u^c}
\newcommand{\dbar}{d^c}
\newcommand{\ebar}{e^c}
\newcommand{\nubar}{\nu^c}
\newcommand{\Ndw}{N_{\rm DW}}
\newcommand{\Fpq}{F_{\rm PQ}}
\newcommand{\fpq}{v_{\rm PQ}}
\newcommand{\Br}{{\rm Br}}
\newcommand{\Lag}{\mathcal{L}}
\newcommand{\Lqcd}{\Lambda_{\rm QCD}}

\newcommand{\ji}{j_{\rm inf}} 
\newcommand{\jb}{j_{B-L}} 
\newcommand{\M}{M} 
\newcommand{\im}{{\rm Im} }
\newcommand{\re}{{\rm Re} }

\def\lrf#1#2{ \left(\frac{#1}{#2}\right)}
\def\lrfp#1#2#3{ \left(\frac{#1}{#2} \right)^{#3}}
\def\lrp#1#2{\left( #1 \right)^{#2}}
\def\REF#1{Ref.~\cite{#1}}
\def\SEC#1{Sec.~\ref{#1}}
\def\FIG#1{Fig.~\ref{#1}}
\def\EQ#1{Eq.~(\ref{#1})}
\def\EQS#1{Eqs.~(\ref{#1})}
\def\TEV#1{10^{#1}{\rm\,TeV}}
\def\GEV#1{10^{#1}{\rm\,GeV}}
\def\MEV#1{10^{#1}{\rm\,MeV}}
\def\KEV#1{10^{#1}{\rm\,keV}}
\def\blue#1{\textcolor{blue}{#1}}
\def\red#1{\textcolor{blue}{#1}}

\newcommand{\eff}{\Delta N_{\rm eff}}
\newcommand{\neff}{\Delta N_{\rm eff}}
\newcommand{\cc}{\Omega_\Lambda}
\newcommand{\Mpc}{\ {\rm Mpc}}
\newcommand{\Msolar}{M_\odot}

\def\my#1{\textcolor{blue}{#1}}
\def\MY#1{\textcolor{blue}{[{\bf MY:} #1}]}


\begin{flushright}
TU-1324\\ 
\end{flushright}

\title{
Reconciling Inflation with the Festina Lente Bound through Affleck-Dine Baryogenesis
}

\author{Kazuto Nakamura and Masaki Yamada}
\affiliation{Department of Physics, Tohoku University, Sendai, Miyagi 980-8578, Japan}

\abstract{
The Festina-Lente bound is a swampland conjecture motivated by the behavior of charged BHs in dS spacetime, which imposes a constraint on the masses of charged particles.
Although relatively weak in the present-day Universe, it can severely constrain inflation.
We clarify its underlying assumptions and show that breaking all gauge symmetries during inflation, with gauge boson masses above the Hubble scale, provides a natural way to evade these constraints.
We catalogue single flat directions in the Minimal Supersymmetric Standard Model that completely break its continuous gauge group, together with pairs that achieve complete breaking only when both directions are excited and simultaneously satisfy F- and D-flatness.
With suitable lifting operators, such condensates can also generate the baryon asymmetry through the Affleck-Dine mechanism.
Within this classification, all single-direction monomials that admit complete breaking carry nonzero $B-L$, providing candidates for Affleck-Dine baryogenesis.
We further show that, in this setting, a large condensate compatible with the observed baryon asymmetry and isocurvature constraints readily satisfies the Festina-Lente bound, establishing a natural connection between the FL bound and Affleck-Dine baryogenesis.
}

\emailAdd{kazuto.nakamura.q7@dc.tohoku.ac.jp}
\emailAdd{m.yamada@tohoku.ac.jp}

\maketitle
\flushbottom

\newpage

\section{Introduction\label{sec:Introduction}}
We do not yet have a complete description of quantum gravity.
Nevertheless, proposed swampland conjectures provide possible consistency conditions for determining which effective field theories can be coupled to quantum gravity~\cite{Vafa:2005ui,Palti:2019pca}.
In this paper, we focus on the Festina Lente (FL) bound, a conjectured lower bound on the masses of charged particles in a de Sitter (dS) background~\cite{Montero:2019ekk,Montero:2021otb}.

The FL bound is motivated by the evaporation of charged Reissner-Nordstr\"{o}m-de Sitter (RN-dS) black holes (BHs).
Rapid discharge can drive a charged Nariai BH outside the parameter region of regular RN-dS solutions and lead to a singular evolution, rather than relaxation toward empty dS space~\cite{Montero:2019ekk}.
Avoiding this behavior is motivated by the thermodynamics of the dS static patch and by the conjecture that its finite entropy reflects a finite-dimensional Hilbert space~\cite{Gibbons:1977mu,Banks:2000fe,Witten:2001kn}.
The FL bound should therefore be regarded as a heuristic consistency condition for this thermodynamic picture, rather than as an established consequence of quantum gravity.
The exclusion of the singular evolution is also closely related to cosmic censorship~\cite{Montero:2019ekk}.
\footnote{
The original proposal did not present direct tests of the FL bound in string theory~\cite{Montero:2019ekk}.
Subsequent work has examined its consistency in string constructions~\cite{Montero:2021otb}, although its general validity remains conjectural.
A backreacted tunneling analysis identified decay channels that can violate the FL bound without producing a crunch for every observer, while leaving their competition with multiparticle discharge unresolved~\cite{Aalsma:2023mkz}.
}

Both inflation and the present dark-energy-dominated era can be approximated by dS or quasi-dS backgrounds.
When its underlying assumptions are satisfied, the FL bound requires every particle of mass $m$ and Abelian charge $q$ to satisfy $m^2 \gtrsim |q|gM_{\mathrm{Pl}}H$, where $g$ is the corresponding gauge coupling and $H$ is the Hubble parameter.
For a non-Abelian gauge theory, massless gauge bosons are themselves charged under the unbroken gauge group.
The FL argument then requires the non-Abelian gauge theory to confine or to be Higgsed above the Hubble scale, parametrically giving $\Lambda_{\rm confinement},m_{\rm gauge}\gtrsim H$~\cite{Montero:2021otb}.
\footnote{
The relation between the confinement and Hubble scales has also been investigated holographically in the Karch-Randall setup~\cite{Mishra:2022fic}.
}
If the Standard Model (SM) particle masses retain their present values during inflation, the electromagnetic FL bound restricts the inflationary energy scale to approximately the MeV scale~\cite{Montero:2019ekk,Lee:2021cor}.
Such a low inflationary scale cannot support reheating above the temperature required by big bang nucleosynthesis~\cite{deSalas:2015glj,Hasegawa:2019jsa}.
The FL bound therefore creates a severe tension with inflation under this assumption.

The FL argument relies on several conditions that also indicate possible ways to avoid this tension.
It assumes four-dimensional semiclassical general relativity throughout the relevant BH evolution.
It also assumes that the quasi-dS phase lasts long enough for rapid discharge and the ensuing geometric evolution to occur.
Finally, the theory must admit charged RN-dS solutions supported by gauge fields that remain effectively long-ranged on the Hubble scale.
One may lower the inflationary scale, alter the charged-particle masses, or make the discharge slower than the duration of inflation~\cite{Montero:2019ekk}.
The last possibility requires an extremely small effective gauge coupling.
An inflaton-dependent gauge kinetic function can generate such a coupling, but it can break the inflaton shift symmetry and produce gauge-field fluctuations~\cite{Montero:2019ekk,Barnaby:2012tk,Jain:2012ga}.

We instead consider removing the long-range gauge fields required by the RN-dS construction.
If all gauge symmetries are Higgsed during inflation and every gauge boson is heavier than the Hubble scale, the corresponding long-range flux cannot support the RN-dS solutions used in the FL argument~\cite{Montero:2021otb}.
The Minimal Supersymmetric Standard Model (MSSM) contains many D- and F-flat directions whose condensates can Higgs gauge symmetries.
These directions can be described by gauge-invariant monomials and have been systematically catalogued~\cite{Gherghetta:1995dv,Basboll:2009tz}.
In this work, we identify individual MSSM flat directions that completely break the gauge group and classify pairs that achieve complete breaking only when both directions are excited.

When scalar condensates Higgs the gauge group, it is natural to ask whether the same condensates also carry and break baryon or lepton number.
If baryon- or lepton-number-violating interactions lift such a direction, the phase motion of the condensate can generate a net baryon or lepton asymmetry through the Affleck-Dine (AD) mechanism~\cite{Affleck:1984fy,Dine:1995uk,Dine:1995kz,Enqvist:2003gh}.
The Higgs solution to the FL argument therefore raises the possibility that the symmetry breaking suggested by the FL bound is connected to the baryon asymmetry of the Universe.
All monomial types for single flat directions identified in our classification not only admit complete gauge symmetry breaking but also naturally carry nonzero $B-L$.
The FL bound thus motivates this class of realizations without determining the baryon asymmetry by itself.
For the AD dynamics, we adopt the $\mathbb Z_4^R$ symmetry motivated by proton stability and study a flavor-aligned single-field limit along $\bar d\bar d\bar dLL$ as an example.
For the fully Higgsed configurations considered here, we also find that the large vacuum expectation value (VEV) required for successful AD baryogenesis readily makes all gauge bosons heavier than the Hubble scale during inflation.

The remainder of this paper is organized as follows.
In Sec.~\ref{sec:FLB_review}, we review the FL argument for charged particles, discuss the duration of the dS phase, and summarize its extension to non-Abelian gauge theories.
In Sec.~\ref{sec:FLB to cosmology}, we apply the FL bound to the present accelerating era and to inflation.
In Sec.~\ref{sec:gauge symmetry breaking}, we identify single flat directions and pairs of directions that completely break the continuous MSSM gauge group.
In Sec.~\ref{sec:ADBG}, we combine the resulting gauge masses with AD dynamics, the baryon abundance, and the isocurvature constraint.
We discuss the implications and limitations of this scenario in Sec.~\ref{sec:discussion}.

\section{Festina Lente (FL) bound\label{sec:FLB_review}}

In this section, we review the FL bound, which is motivated by requiring charged BHs in dS spacetime to evaporate without driving the static patch into a singular configuration~\cite{Montero:2019ekk,Montero:2021otb}.

In Sec.~\ref{subsec:RN-dS BH}, we describe charged BHs in dS spacetime and identify the parameter region in which their singularities are hidden behind horizons.
In Sec.~\ref{sec:FLB}, we introduce the FL bound by requiring Schwinger discharge of Nariai BHs to be suppressed.
In Sec.~\ref{sec:duration}, we examine the role of the duration of the dS phase by comparing it with the discharge timescale.
In Sec.~\ref{sec:YM}, we extend the discussion to non-Abelian gauge theories and describe how confinement or the Higgs mechanism can remove the long-range gauge fields required by the argument.
Finally, in Sec.~\ref{sec:conditions}, we summarize the conditions and assumptions underlying the FL bound.

\subsection{The family of Reissner-Nordstr\"{o}m-de Sitter black holes\label{subsec:RN-dS BH}}

The charged BH solution of interest in dS spacetime is the RN-dS BH~\cite{Romans:1991nq}.
This solution can be obtained from the following action for the (3+1)-dimensional Einstein-Maxwell-de Sitter system with metric signature $(-,+,+,+)$
\begin{align}
S = \int \dd^4 x \sqrt{-g} \left[
\frac{1}{16\pi G} \left( R - \frac{6}{\ell^2} \right)
- \frac{1}{4 g^2} F_{\mu\nu} F^{\mu\nu}
\right],
\end{align}
where $\ell$ is the dS radius, defined by $\ell\equiv 1/H$.
The cosmological constant $\Lambda$ is related to $H$ by
\begin{align}
    \Lambda=3H^2.
\end{align}
The RN-dS solution has the metric
\begin{align}
\dd s^2 &= -U(r)\,\dd t^2 + \frac{\dd r^2}{U(r)} + r^2 \dd\Omega_2^2, \\[10pt]
U(r) &\equiv 1 - \frac{2G M_r}{r} + \frac{G (gQ_r)^2}{4\pi r^2} - r^2H^2 
\\
&= 1-\frac{2M}{Hr}+\frac{Q^2}{(Hr)^2}-(Hr)^2, 
\end{align}
where $M_r$ is the BH mass and $Q_r$ is its integer-quantized charge.
The non-negative dimensionless parameters are defined by 
\begin{align}
    M\equiv GM_rH,\; Q^2\equiv \frac{G(gQ_r)^2}{4\pi}H^2.\label{eq:regularization}
\end{align}
The BH produces a static electric field, with gauge potential and field strength given by
\begin{align}
\begin{aligned}
    &A= \frac{g^2}{4\pi}\frac{Q_r}{r} \dd t,\\[5pt]
    &F= -\frac{g^2}{4\pi}\frac{Q_r}{r^2} \,\dd r\wedge \dd t.
\end{aligned}
\end{align}

In general, this system has three horizons, the Cauchy horizon $r_-$, the event horizon $r_+$, and the cosmological horizon $r_c$, with $r_- < r_+ < r_c$.
\begin{itemize}
    \item Cauchy horizon $r_-$:
    The Cauchy horizon is the boundary of the region where the future evolution of spacetime is uniquely determined from given initial data.
    Beyond this horizon, the initial data are no longer sufficient to predict the future uniquely, and the deterministic nature of general relativity breaks down.
    \item Event horizon $r_+$:
    Around a BH, the gravitational field becomes stronger toward the center.
    Inside a certain radius, even light cannot escape to the exterior.
    This boundary is called the event horizon.
    \item Cosmological horizon $r_c$:
    In dS spacetime, sufficiently distant events cannot send signals that ever reach a given observer.
    This boundary is called the cosmological horizon.
\end{itemize}
Two of these horizons, the event horizon at $r_+$ and the cosmological horizon at $r_c$, are accessible to an observer outside the BH.
The locations of these horizons are constrained by the requirement that the solution describe a BH without naked singularities.
The existence of an event horizon is therefore crucial for determining whether the singularity is naked.
If $U(r)=0$ has no real solution corresponding to the event horizon $r_+$, the singularity at $r=0$ is not hidden.
To assess this condition, we examine the discriminant $\Delta$ of $U(r)=0$,
\begin{align}
\Delta \equiv M^2 - Q^2 - 27 M^4 + 36 M^2 Q^2 - 8 Q^4 - 16 Q^6.
\end{align}
\begin{figure}
  \centering
  \vspace*{0.2cm}
  \includegraphics[width=0.5\linewidth]{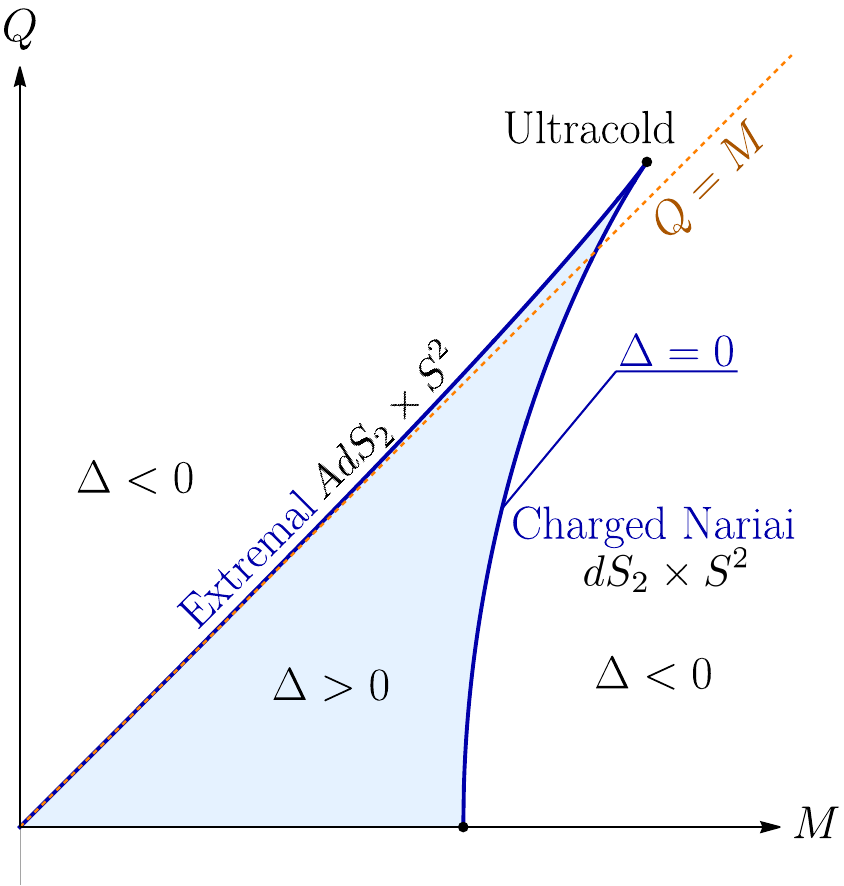}\\
  \vspace{0.7cm}
  \caption{ 
  Shark-fin region with $\Delta \geq 0$ (blue).
  The axes show the dimensionless mass $M$ and charge $Q$ defined in Eq.~(\ref{eq:regularization}).
  The FL bound assumes that the RN-dS BH remains within this region throughout its evolution.}
  \label{fig:shark fin}
\end{figure}

If $\Delta>0$, there are no naked singularities, while $\Delta=0$ corresponds to a boundary consisting of two different branches~\cite{Romans:1991nq}.
\begin{itemize}
    \item Extremal branch ($r_-=r_+$):
    In this limit, the BH has zero temperature with a horizon smaller than the cosmological horizon, and the inner and outer BH horizons coincide, $r_-=r_+$.
    The near-horizon geometry is $AdS_2\times S^2$.
    \item Charged Nariai branch ($r_+=r_c$):
    In this limit, the event horizon and the cosmological horizon are in thermal equilibrium, sharing the same Hawking temperature.
    Their apparent coincidence is a coordinate effect.
    Away from the ultracold endpoint, an appropriate rescaling gives the $dS_2\times S^2$ geometry, in which the two horizons have a finite proper separation~\cite{Montero:2019ekk,Castro:2022cuo}.
    At the ultracold endpoint, all three horizons coincide and the two-dimensional factor becomes flat~\cite{Castro:2022cuo}.
\end{itemize}
The parameter region defined by $\Delta>0$ is shown in \FIG{fig:shark fin} and is referred to as the ``shark fin.''
Within this region, the static RN-dS solutions possess BH horizons that hide the central singularity.

However, a problem arises when we consider the evaporation of RN-dS BHs.
Even if a BH initially lies inside the shark-fin region, it can leave this region during evaporation.
The static patch can then evolve into a singular configuration without returning to empty dS space.
In particular, in the context of the FL bound, we focus on the right boundary of the shark fin, which corresponds to the charged Nariai branch.
The BHs on this boundary are called charged Nariai BHs.

\subsection{Festina Lente Bound}
\label{sec:FLB}

The FL bound is motivated by requiring the evaporation of charged Nariai BHs to be compatible with the thermal picture of the dS static patch~\cite{Montero:2019ekk,Montero:2021otb}.
At fixed positive cosmological constant, the total horizon entropy of an RN-dS BH satisfies
\begin{align}
    S_c+S_{\rm BH}=\frac{\pi}{G}(r_c^2+r_+^2)
    <\frac{\pi}{GH^2}=S_{\rm dS}^{(0)},
    \label{eq:RN-dS entropy bound}
\end{align}
where $S_c$ and $S_{\rm BH}$ are the cosmological and BH horizon entropies, and $S_{\rm dS}^{(0)}$ is the entropy of empty dS space~\cite{Gibbons:1977mu}.
This comparison supports the expectation that the equilibrium endpoint is empty dS space. 
However, by itself, it does not determine the evaporation dynamics or rule out all configurations outside the shark-fin region. 
The FL argument imposes the additional requirement that discharge should not drive the entire static patch into a singular evolution instead of allowing relaxation toward empty dS space.

Hawking radiation~\cite{Hawking:1975vcx} and Schwinger pair production~\cite{Schwinger:1951nm} govern the semiclassical evaporation of charged BHs.
We summarize the relevant semiclassical results of Ref.~\cite{Montero:2019ekk} below and refer to that work for the evolution equations and their derivation.

The $dS_2\times S^2$ geometry of the Nariai limit allows curvature effects on pair production to be studied using the Schwinger effect in two-dimensional dS space~\cite{Frob:2014zka}.
In the strong-field limit, the Nariai current reproduces the flat-space Schwinger result~\cite{Montero:2019ekk}.
In this regime, the pair production rate has the exponential dependence
\begin{align}
    \Gamma\propto\exp\left(-\frac{\pi m^2}{\abs{qE}}\right),
    \label{eq:schwinger pair production}
\end{align}
where $m$ and $q$ are the mass and charge of the produced particle and $E$ is the electric field.
The prefactor depends on the particle species and spacetime geometry.
In the quasistatic regime, $m^2\gg\abs{qE}$ for the relevant charged particles, and the Schwinger factor suppresses the current.
The geometry can then be approximated by a sequence of RN-dS solutions whose mass and charge evolve slowly through Hawking radiation and Schwinger pair production.
Within the semiclassical treatment, the resulting trajectories are compatible with relaxation toward empty dS space~\cite{Montero:2019ekk}.

If a sufficiently light charged species makes pair production unsuppressed, the electric field can instead be screened rapidly.
For semiclassical Nariai BHs, Hawking radiation changes the geometry slowly, whereas unsuppressed Schwinger pair production can rapidly screen the electric field in the strong-field regime.
We therefore neglect the backreaction of Hawking radiation during this rapid discharge~\cite{Montero:2019ekk}.
Then its energy is transferred to charged particles and radiation.
In this adiabatic regime, the screened field is replaced by radiation to follow the subsequent geometric evolution.
For the charged Nariai initial conditions considered there, the geometry initially takes the form $dS_2\times S^2$, and the subsequent collapse of the two-sphere drives the entire static patch into a crunch.
Avoiding this evolution motivates the condition
\begin{align}
    m^2\gtrsim\abs{qE}
\end{align}
for every charged particle, up to numerical factors of order unity.

On the Nariai branch, the maximum electric field is
\begin{align}
    E_{\rm max}=\sqrt{6}gM_{\mathrm{Pl}}H,
\end{align}
where $M_{\mathrm{Pl}}=(8\pi G)^{-1/2}$ is the reduced Planck mass and $g$ is the gauge coupling.
Requiring the condition to hold over the Nariai branch gives the FL bound~\cite{Montero:2021otb},
\begin{align}
    m^2\gtrsim\abs{q}g M_{\mathrm{Pl}} H\quad\text{for every charged particle},
    \label{FLB}
\end{align}
where we omit the numerical factor of order unity.

\subsection{Duration of dS spacetime}
\label{sec:duration}

The FL argument also requires the approximately dS background to persist over the relevant discharge and geometric evolution timescales.
If inflation ends before the singular evolution can occur, the argument must be reconsidered in the subsequent background.
We denote the duration of inflation and the discharge timescale by $\Delta t_{\rm inf}$ and $\Delta t_{\rm disch}$, respectively.
The estimates used below are
\begin{align}
    \Delta t_{\rm inf}&\sim\frac{N_e}{H},\label{eq: inflation duration}\\
    \Delta t_{\rm disch}&\sim\frac{1}{\sqrt{\abs{qE}}},\label{eq: discharge duration}
\end{align}
where $N_e$ is the number of e-folds. 
The second estimate applies in the strong-field regime $\abs{qE}\gg m^2,H^2$, 
in which the mass and curvature scales can be neglected in the local pair production process~\cite{Montero:2019ekk}.

Formally imposing $\Delta t_{\rm disch}>\Delta t_{\rm inf}$ gives
\begin{align}
    g\lesssim\frac{H}{N_e^2 M_{\mathrm{Pl}}}\simeq1.2\times10^{-9}\left(\frac{H}{10^{13}\GeV}\right)\left(\frac{60}{N_e}\right)^2,
    \label{eq:constraintfore-folding}
\end{align}
where we have taken $\abs{q}=1$ and used $E\sim g M_{\mathrm{Pl}} H$.

The above estimate for $\Delta t_{\rm disch}$ assumes the strong-field regime $|qE|\gg H^2$, in which discharge occurs within a Hubble time.
Requiring $\Delta t_{\rm disch}>N_e/H$ instead places the system outside this regime, so the resulting bound on $g$ is an extrapolation beyond the validity of the estimate. 
We therefore treat Eq.~(\ref{eq:constraintfore-folding}) only as a formal estimate when discussing the finite duration of inflation.
Establishing such a criterion requires control of the discharge dynamics with dS expansion and of the duration of the background.

\subsection{Generalization to non-Abelian gauge theories}
\label{sec:YM}

The FL bound can be generalized to non-Abelian gauge theories~\cite{Montero:2021otb}.
We consider QCD as an example, but the same argument applies to other non-Abelian gauge theories.

More explicitly, we consider the following configuration, in which only the fields along a particular Cartan direction are nonzero and all other fields are set to zero
\begin{align}
    A_\mu=A_\mu^{(C)}T^C 
    \\
    T^C = n_3 T^3+n_8 T^8
\end{align}
where $T^3$ and $T^8$ are the Cartan generators of the Lie algebra of $\mathfrak{su}(3)$, and $n_3$ and $n_8$ are constants. 
Since these two generators commute, the non-Abelian field strength tensor can take the following Maxwell-like form
\begin{align}
    F_{\mu\nu}&=F_{\mu\nu}^{(C)} T^C 
\end{align}
where $F_{\mu\nu}^{(C)}$ is the field strength that supports RN-dS BHs, as in the $U(1)$ case.
Therefore, the RN-dS BH discussed in Sec.~\ref{subsec:RN-dS BH} can be constructed using an $SU(3)_c$ gauge field.
Moreover, gluons carry color charge, and the relevant charged fields in this setup include, for example, $A^{\pm}_{\mu}=A^1_\mu\pm iA^2_\mu $.
Thus, the massless color-charged gluons of $SU(3)_c$ violate the FL bound in Eq.~(\ref{FLB}).

We can avoid this violation if the confinement scale of the $SU(3)_c$ gauge theory exceeds the temperature of the dS static patch,
\begin{align}
    \Lambda_{\rm confinement}\gtrsim H.\label{confinement FLB}
\end{align}
The right-hand side of Eq.~(\ref{confinement FLB}) is of the order of the dS temperature $T_{\rm dS}=H/2\pi$~\cite{Gibbons:1977mu}.
For this confinement solution, the confinement scale of the relevant non-Abelian sector must exceed the Hubble parameter.
If this inequality is satisfied, a non-Abelian gauge field cannot support an RN-dS BH with a long-range flux, even when the field is restricted to the Cartan directions. 

Another possible way to evade the FL bound is for all gauge fields to acquire masses through the Higgs mechanism.
In this case, the gauge field cannot support a $1/r$ field profile.
Consequently, while the Higgs phase is maintained, the long-range flux required by the RN-dS construction is absent on the Hubble scale.
The condition for evading the FL bound through the Higgs mechanism is~\cite{Montero:2021otb}
\begin{align}
    m_{\rm gauge}\gtrsim H,\label{eq: FLB for gauge mass}
\end{align}
where $m_{\rm gauge}$ denotes the gauge boson mass. 
This condition will play an important role in the following discussion.

\subsection{Summary of the assumptions underlying the FL bound}
\label{sec:conditions}

Here we summarize the assumptions underlying the FL bound~\cite{Montero:2019ekk,Montero:2021otb}.
The FL bound applies when the following assumptions about the background spacetime and the particle physics model hold.

\def\theenumi{\roman{enumi}}
\def\labelenumi{\theenumi)}

We make the following assumptions.
\begin{enumerate}
\item The background spacetime is four-dimensional, semiclassical, and quasi-dS, with a nearly constant Hubble parameter $H$ over the relevant timescale.
\item The lifetime of the system exceeds the timescale for the catastrophic evolution of a Nariai BH.
\item A charged Nariai solution exists for a gauge field whose range exceeds the Hubble radius.
\item The evolution of a Nariai BH must not lead to a high-curvature region encompassing the entire static patch.
\end{enumerate}

\def\theenumi{\alph{enumi}}
\def\labelenumi{\theenumi)}

The FL bound imposes the following conditions on the gauge sectors. 
\begin{enumerate}
\item For an unbroken Abelian gauge symmetry, a particle with mass $m$ and charge $q$ must satisfy $m^2 \gtrsim  \abs{q} g \Mpl H$.
Alternatively, the gauge symmetry must be Higgsed above the Hubble scale. 
\item A non-Abelian gauge sector must be confined or Higgsed above the Hubble scale.%
\footnote{Partial breaking can also be consistent if any residual non-Abelian factors are confined above the Hubble scale and all states charged under surviving Abelian factors satisfy the corresponding FL mass bound. 
}
\end{enumerate}

If any of these assumptions is violated, this argument does not imply catastrophic evolution. 
In particular, the quasi-dS phase may last for a sufficiently small number of e-folds. 
The argument also does not apply when no gauge field remains effectively long-ranged on the Hubble scale, for example because the gauge sector is confined or fully Higgsed. 
In the next section, we compare these possibilities in cosmological settings and motivate gauge symmetry breaking during inflation.

\section{Application of the FL bound to cosmology\label{sec:FLB to cosmology}}

We now apply the conditions established above to the present dark energy era and to inflation.
The accelerated expansion in these epochs may be driven by a cosmological constant or by a dynamical field, such as quintessence or the inflaton.
In each case, the application of the FL bound requires an approximately dS background that persists over the relevant BH evolution timescale.
We first examine the present SM vacuum and then discuss how the larger Hubble scale during inflation constrains the particle spectrum and gauge sectors.

\subsection{Dark energy}

We first consider the case in which dark energy is a positive cosmological constant and the Universe approaches dS spacetime.
We take $H_{\rm cc}\sim 10^{-33}\eV$ as the characteristic Hubble scale of this phase and assume that the lifetime condition in Sec.~\ref{sec:conditions} is satisfied.
In the present SM vacuum, the unbroken electromagnetic gauge symmetry is U(1)$_e$, and the electron is the lightest electrically charged particle.
Setting $g=e$ and $\abs{q}=1$ in Eq.~(\ref{FLB}), where $e\simeq 0.3$ is the electromagnetic coupling, we obtain
\begin{align}
    m_e\gtrsim \sqrt{e M_{\mathrm{Pl}} H_{\rm cc}}\sim 10^{-3}\eV.
    \label{eq:FLB electron today}
\end{align}
The electron mass, $m_e\simeq 0.511\MeV$, exceeds this lower bound by more than eight orders of magnitude.
The other electrically charged SM particles also satisfy the bound, whereas electrically neutral particles are not constrained by this electromagnetic bound.

For QCD, the confinement scale is $\Lambda_{\rm QCD}\approx 200\MeV$, which is much larger than $H_{\rm cc}$.
The condition in Eq.~(\ref{confinement FLB}) is therefore satisfied.
As discussed in Sec.~\ref{sec:YM}, confinement prevents the color gauge fields from supporting the long-range flux required for an RN-dS BH.
The weak gauge sector instead satisfies the condition through the Higgs mechanism.
The massive $W$ and $Z$ bosons have masses of order $10^2\GeV\gg H_{\rm cc}$, so the corresponding fields satisfy Eq.~(\ref{eq: FLB for gauge mass}) and cannot support the required long-range fields.
Thus, the SM spectrum is consistent with the FL bound at the present dark energy scale, with electromagnetism satisfying the charged-particle mass bound and the other gauge fields subject to confinement or the Higgs mechanism~\cite{Montero:2019ekk,Montero:2021otb}.

If dark energy is generated by a rolling scalar field such as quintessence, the application of the FL bound is conditional on the assumptions in Sec.~\ref{sec:conditions}~\cite{Montero:2021otb}.
In particular, the background must remain approximately dS over the relevant BH evolution timescale, and the accelerating phase must last sufficiently long.
In any cases, the present SM spectrum satisfies the same mass and confinement conditions.

\subsection{Inflation}
\label{sec:inflation}

We now apply the conditions in Sec.~\ref{sec:conditions} to a quasi-dS inflationary phase with Hubble parameter $H_{\rm inf}$.
One possible way to evade the FL bound is to make the discharge timescale longer than the duration of inflation, as discussed in Sec.~\ref{sec:duration}.
The estimate in Eq.~(\ref{eq:constraintfore-folding}) suggests that this requires a very small gauge coupling.
An inflaton-dependent gauge kinetic function could keep the gauge coupling sufficiently small during inflation~\cite{Montero:2019ekk}.
However, such a coupling requires care to preserve the flatness of the inflaton potential.
Depending on its form, it can break the inflaton shift symmetry or induce gauge field production, potentially generating non-Gaussianity or correlations between curvature perturbations and magnetic fields~\cite{Barnaby:2012tk,Jain:2012ga}.

If the inflationary phase lasts long enough for the FL argument to apply, the QCD condition in Eq.~(\ref{confinement FLB}) gives
\begin{align}
    \Lambda_{\rm QCD}\gtrsim H_{\rm inf}.
\end{align}
Here, $\Lambda_{\rm QCD}$ must be evaluated during inflation.
If it remains close to its present value, $\Lambda_{\rm QCD}\approx 200\MeV$, this condition places a stringent upper bound on the inflationary scale.
For an energy density $\rho_{\rm inf}$ dominated by the inflaton potential, the Friedmann equation gives the parametric relation
\begin{align}
    E_{\rm inf}\equiv \rho_{\rm inf}^{1/4}\sim \sqrt{ M_{\mathrm{Pl}} H_{\rm inf}}.
    \label{eq:inflation energy scale}
\end{align}
The confinement condition then implies%
\footnote{
Possible extensions of the FL argument to finite-temperature backgrounds have been discussed in Ref.~\cite{Venken:2023hfa}.
}
\begin{align}
    10^9\GeV\gtrsim E_{\rm inf}.
\end{align}
This bound excludes high-scale inflation.
The constraint can be satisfied by sufficiently low-scale inflation or by an enhanced confinement scale during inflation~\cite{Jeong:2013xta,Takahashi:2015waa}.

Confinement of QCD alone does not ensure consistency with the FL bound.
Every other non-Abelian gauge sector must also satisfy the confinement or Higgs condition in Sec.~\ref{sec:YM}.
Moreover, if an Abelian gauge field remains long-ranged, every particle charged under it must satisfy Eq.~(\ref{FLB}), with its mass and gauge coupling evaluated during inflation.
In particular, breaking the electroweak symmetry while leaving electromagnetism unbroken does not remove the mass bound on electrically charged particles~\cite{Montero:2021otb,Lee:2021cor}.
Again, setting $g=e = \mathcal{O}(1)$ and $\abs{q}=1$ in Eq.~(\ref{FLB}), we obtain
\begin{align}
    H_{\rm inf} \lesssim \frac{m_e^2}{e M_{\mathrm{Pl}}}  \sim 10^{-24}\GeV
\end{align}
or 
\begin{align}
    E_{\rm inf} \lesssim \frac{m_e}{e^{1/2}}  \sim \MeV
\end{align}
where we assume that the electron mass during inflation is also given by $m_e \sim \mathcal{O}(1) \MeV$. 
The maximal reheating temperature is then below the few-MeV temperature required by nucleosynthesis and neutrino thermalization~\cite{deSalas:2015glj}.

Another way to evade the FL bound is to break all gauge symmetries spontaneously during inflation, as we discuss in the next section.
The resulting gauge boson masses must satisfy Eq.~(\ref{eq: FLB for gauge mass}), with $H=H_{\rm inf}$, so that no gauge field remains effectively long-ranged on the Hubble scale.
This possibility removes the charged Nariai solutions required by the FL argument without relying on an exceptionally small gauge coupling or a confinement scale above the inflationary Hubble scale.

\section{Gauge symmetry breaking along supersymmetric flat directions \label{sec:gauge symmetry breaking}}
\label{sec:gauge breaking}

The discussion in Sec.~\ref{sec:YM} identifies Higgsing as a way to remove the long-range gauge fields required by the FL argument.
We consider scalar condensates that break all continuous MSSM gauge symmetries during inflation and give every gauge boson a mass much larger than $H_{\rm inf}$.
Supersymmetric flat directions allow such expectation values during inflation, and the subsequent decay of the condensates allows the fields to approach the ordinary MSSM vacuum.

In Sec.~\ref{sec:flat directions}, we identify flat directions along which a single condensate completely breaks the continuous gauge group.
In Sec.~\ref{sec:two flat directions}, we classify pairs that achieve complete breaking only when both condensates are present.

\subsection{Complete gauge symmetry breaking along a single flat direction}
\label{sec:flat directions}

Supersymmetric theories can admit flat directions along which the scalar potential vanishes in the supersymmetric limit at the renormalizable level.
To illustrate how such directions arise, we briefly review the scalar potential in global supersymmetry.
For canonical kinetic terms, the scalar potential is a sum of F-term and D-term contributions~\cite{Gherghetta:1995dv},
\begin{align}
    V=V_F+V_D,\qquad
    V_F=\sum_i\left|\frac{\partial W}{\partial\phi_i}\right|^2,\qquad
    V_D=\frac12\sum_A g_A^2\left(\sum_i\phi_i^\dagger T_i^A\phi_i\right)^2,
    \label{eq:global SUSY potential}
\end{align}
where $W$ is the superpotential, $\phi_i$ are scalar components of chiral superfields, and $T_i^A$ are the Hermitian generators in their respective representations.
We assume that there is no Fayet-Iliopoulos term.
A direction is flat in this limit if all F-terms and D-terms vanish along it.

To illustrate how D-flatness can coexist with complete gauge symmetry breaking, we first neglect supergravity effects and consider an $SU(N)$ theory with vanishing superpotential.
Suppose that $F$ ($\geq N$) chiral multiplets transform in the fundamental representation, with scalar components $\phi_{a i}$, where $a=1,\ldots,N$ and $i=1,\ldots,F$ are gauge and flavor indices.
For the $N\times F$ matrix $\phi$, consider the configuration
\begin{align}
    \langle\phi\rangle=v\begin{pmatrix}\mathbf{1}_N&\mathbf{0}_{N\times(F-N)}\end{pmatrix},\qquad
    \sum_{i=1}^{F}\langle\phi_i\rangle\langle\phi_i\rangle^\dagger=|v|^2\mathbf{1}_N.
    \label{eq:SU N flat configuration}
\end{align}
Since every $SU(N)$ generator is traceless, this configuration satisfies
\begin{align}
    \sum_i\langle\phi_i\rangle^\dagger T^A\langle\phi_i\rangle
    =|v|^2\operatorname{Tr}T^A=0,
    \label{eq:SU N D flatness}
\end{align}
and hence $V_D=0$ for arbitrary $v$.
For $v\neq0$, the matrix has rank $N$, and the only gauge transformation satisfying $U\langle\phi\rangle=\langle\phi\rangle$ is $U=\mathbf{1}_N$.
Thus, the gauge symmetry is completely broken even though the scalar potential is flat.
More generally, gauge-invariant monomials describe D-flat directions, but their constituent expectation values must still satisfy the D-flatness conditions~\cite{Luty:1995sd,Gherghetta:1995dv}.

In the MSSM, D-flatness must hold simultaneously for $SU(3)_c$, $SU(2)_L$, and $U(1)_Y$.
F-flatness additionally requires the derivatives of the renormalizable MSSM superpotential to vanish.
From the MSSM monomials catalogued in Ref.~\cite{Gherghetta:1995dv}, Table~\ref{tab: Flat direction} lists those associated with F- and D-flat directions that can break the full continuous gauge group, with gauge and flavor contractions left implicit.
Here, $Q$ and $L$ are quark and lepton doublets, while $\bar u$, $\bar d$, and $\bar e$ are the charge-conjugate singlet superfields.
The subscript $4$ denotes the weak-isospin quartet obtained from three $Q$ doublets.
The monomial specifies the field content, while the breaking pattern depends on the actual expectation values.

For example, a D-flat configuration associated with $\bar d\bar d\bar dLL$ is
\begin{align}
    &\langle\bar d_1\rangle=\frac{v}{\sqrt{5}}\begin{pmatrix}1\\0\\0\end{pmatrix},\quad
    \langle\bar d_2\rangle=\frac{v}{\sqrt{5}}\begin{pmatrix}0\\1\\0\end{pmatrix},\quad
    \langle\bar d_3\rangle=\frac{v}{\sqrt{5}}\begin{pmatrix}0\\0\\1\end{pmatrix},\notag\\[3pt]
    &\langle L_i\rangle=\frac{v}{\sqrt{5}}\begin{pmatrix}1\\0\end{pmatrix},\quad
    \langle L_j\rangle=\frac{v}{\sqrt{5}}\begin{pmatrix}0\\1\end{pmatrix},\qquad i\neq j,
    \label{eq:dddLL configuration}
\end{align}
with all other expectation values set to zero. 
The factor $1/\sqrt{5}$ normalizes the sum of the five nonzero squared amplitudes to $|v|^2$, so that $v$ is the expectation value of the canonically normalized complex field along this direction.
The three color contributions and the two weak-isospin contributions each sum to a matrix proportional to the identity.
Their non-Abelian D-terms therefore vanish, while the hypercharge contribution cancels because $3Y_{\bar d}+2Y_L=3(1/3)+2(-1/2)=0$.
The renormalizable F-terms vanish for this configuration in the MSSM with conserved R-parity.
The three independent $\bar d$ expectation values leave no continuous combination of color and hypercharge unbroken, and the two independent $L$ expectation values also break $SU(2)_L$ completely.
This provides an explicit realization of the symmetry breaking required in Sec.~\ref{sec:YM}.

\begin{table}[t]
    \centering
    \begin{tabular}{ccccc}
        \toprule
        F- and D-flat direction & $B$ & $L$ & $B-L$ & $n_I=2\deg I$ \\
        \midrule
        $\bar{d}\bar{d}\bar{d}LL$ & $-1$ & +2 & $-3$ & 10 \\
        $Q\bar{u}Q\bar{u}\bar{e}$ & 0 & $-1$ & +1 & 10 \\
        $QQQQ\bar{u}$ & +1 & 0 & +1 & 10 \\
        $(QQQ)_4LLL\bar{e}$ & +1 & +2 & $-1$ & 14 \\
        $\bar{u}\bar{u}\bar{d}Q\bar{d}Q\bar{d}$ & $-1$ & $0$ & $-1$ & 14\\
        \bottomrule
    \end{tabular}
    \caption{Gauge-invariant monomials $I$ corresponding to renormalizable F- and D-flat directions in the MSSM that can break the full continuous gauge group.
Gauge and flavor contractions are implicit.
The corresponding charges $B$, $L$, and $B-L$ are also listed~\cite{Gherghetta:1995dv}.
The last column gives the degree of $I^2$, which is allowed by the $\mathbb Z_4^R$ symmetry adopted in Sec.~\ref{sec:AD dynamics}.
}
    \label{tab: Flat direction}
\end{table}

The potential is lifted by supersymmetry breaking and higher-dimensional operators.
During inflation, supergravity interactions can generate a negative mass squared of order $H_{\rm inf}^2$ along a flat direction and drive it away from the origin~\cite{Dine:1995uk,Dine:1995kz}.
Higher-dimensional terms can stabilize the field at a large but finite value.
For the present application, every gauge boson mass must satisfy Eq.~(\ref{eq: FLB for gauge mass}) during inflation.
Since these masses scale with the gauge couplings and the scalar expectation values, this condition constrains the magnitude as well as the orientation of the condensate.

As shown in Table~\ref{tab: Flat direction}, all these flat directions carry nonzero $B-L$.
With suitable charge-violating interactions, their condensates can generate a $B-L$ asymmetry through the Affleck-Dine mechanism~\cite{Affleck:1984fy,Dine:1995kz}.
The same condensate can therefore break the gauge symmetries during inflation and later participate in baryogenesis, as we will discuss in Sec.~\ref{sec:AD baryogenesis}.
We emphasize that requiring a single flat direction to break all gauge symmetries yields directions that naturally carry nonzero $B-L$.
This property is important for generating the observed baryon asymmetry, since electroweak sphaleron processes relate the baryon asymmetry after sphaleron processing to the initial $B-L$ asymmetry as $B_{\rm now}=c_{\rm sph}(B-L)$, where $c_{\rm sph}$ is determined by the particle content and the relevant equilibrium conditions~\cite{Harvey:1990qw}.

\subsection{Complete gauge symmetry breaking along two flat directions}
\label{sec:two flat directions}

Complete gauge symmetry breaking can also arise from two condensates, each of which leaves a continuous gauge subgroup unbroken when excited alone.
We classify such pairs using the basic MSSM monomial types of Refs.~\cite{Gherghetta:1995dv,Basboll:2009tz}.
We exclude the five types in Table~\ref{tab: Flat direction}, which already admit complete breaking individually.
The matter-field types considered here are
\begin{align}
    \mathcal C_{\rm m}=\{
    LL\bar e,\ \bar u\bar d\bar d,\ Q\bar dL,\ QQQL,\
    Q\bar uQ\bar d,\ \bar u\bar u\bar d\bar e,\ Q\bar uL\bar e,\
    \bar u\bar u\bar u\bar e\bar e\}.
    \label{eq:pair matter types}
\end{align}
For completeness, we also consider the Higgs types $LH_u$ and $H_uH_d$ with the electroweak-scale $\mu$ term neglected when testing their flatness.
We denote the resulting ten types by $\mathcal C=\mathcal C_{\rm m}\cup\{LH_u,H_uH_d\}$.

The renormalizable superpotential is~\cite{Gherghetta:1995dv}
\begin{align}
    W_{\rm ren}=\mu H_uH_d
    +(Y_u)_{ij}H_uQ_i\bar u_j
    +(Y_d)_{ij}H_dQ_i\bar d_j
    +(Y_e)_{ij}H_dL_i\bar e_j,
    \label{eq:MSSM pair superpotential}
\end{align}
where gauge contractions are implicit and repeated indices are summed.
For configurations with $H_u=H_d=0$, the nontrivial F-flatness conditions reduce to
\begin{align}
    \sum_{i,j,a}(Y_u)_{ij}Q_i^{a\alpha}\bar u_{ja}&=0,\notag\\
    \sum_{i,j,a}(Y_d)_{ij}Q_i^{a\alpha}\bar d_{ja}
    +\sum_{i,j}(Y_e)_{ij}L_i^\alpha\bar e_j&=0,
    \qquad \alpha=1,2.
    \label{eq:pair F flatness}
\end{align}
These conditions must hold with both condensates present, since fields from the two directions can contribute to the same Higgs F-term~\cite{Allahverdi:2006xh}.
The D-terms in Eq.~(\ref{eq:global SUSY potential}) must likewise vanish on the combined configuration.

For canonical scalar and gauge kinetic terms, the gauge boson mass matrix is
\begin{align}
    (M_V^2)_{AB}=g_Ag_B\sum_i
    \langle\phi_i\rangle^\dagger\{T_i^A,T_i^B\}\langle\phi_i\rangle.
    \label{eq:AD gauge mass matrix}
\end{align}
A pair qualifies if there is a family of simultaneous D-flat and F-flat configurations with two independently variable amplitudes, for which this matrix has rank 12 when both amplitudes are nonzero.
Setting either amplitude to zero must recover a configuration with a smaller rank.
This criterion includes gauge generators that mix hypercharge with diagonal color or weak generators.

Table~\ref{tab:two flat directions} gives the resulting 27 pairs of monomial types.
Pairs are unordered, and two different flavor realizations of the same type are included.
Four pairs use two different realizations of the same type, leaving 23 pairs of distinct types.
For each allowed pair, the flavor vectors can be chosen to satisfy Eq.~(\ref{eq:pair F flatness}) while maintaining D-flatness for independently variable amplitudes.
For example, conditions of the form $q^TY_u u=0$ and $q^TY_d d=0$ can be imposed by choosing vectors in the corresponding flavor null spaces.
Here, $q,u,d$ denote flavor vectors.
The $B-L$ charges of the two selected monomials alone do not determine whether the joint configuration can generate a $B-L$ asymmetry.
Even when both are neutral, other invariants involving fields from both directions can carry nonzero $B-L$.
For example, a simultaneous $QQQL+\bar u\bar u\bar d\bar e$ configuration can have $Q\bar dL\ne0$.
Charge generation must therefore be assessed from the physical phase directions and charge-violating operators on the joint flat locus.

\begin{table}[t]
    \centering
    \small
    \begin{tabular}{lll}
        \toprule
        $B-L$ of first type & First type & Allowed second types  \\
        \midrule
       $-1$ & $LL\bar e$ & $\bar u\bar d\bar d$, $QQQL$, $Q\bar uQ\bar d$, $\bar u\bar u\bar d\bar e$, $\bar u\bar u\bar u\bar e\bar e$  \\
        $-1$ & $\bar u\bar d\bar d$ & $Q\bar dL$, $QQQL$, $Q\bar uQ\bar d$, $Q\bar uL\bar e$  \\
        $-1$ & $Q\bar dL$ & $Q\bar dL$, $QQQL$, $Q\bar uQ\bar d$, $\bar u\bar u\bar d\bar e$, $Q\bar uL\bar e$, $\bar u\bar u\bar u\bar e\bar e$  \\
        $0$ & $QQQL$ & $QQQL$, $Q\bar uQ\bar d$, $\bar u\bar u\bar d\bar e$, $Q\bar uL\bar e$, $\bar u\bar u\bar u\bar e\bar e$  \\
        $0$ & $Q\bar uQ\bar d$ & $Q\bar uQ\bar d$, $\bar u\bar u\bar d\bar e$, $Q\bar uL\bar e$, $\bar u\bar u\bar u\bar e\bar e$  \\
        $0$ & $\bar u\bar u\bar d\bar e$ & $Q\bar uL\bar e$  \\
        $0$ & $Q\bar uL\bar e$ & $Q\bar uL\bar e$, $\bar u\bar u\bar u\bar e\bar e$  \\
        \bottomrule
    \end{tabular}
    \caption{Unordered pairs of basic MSSM monomial types admitting simultaneous renormalizable F- and D-flat configurations that completely break the continuous gauge group when both amplitudes are nonzero.
}
    \label{tab:two flat directions}
\end{table}

In the explicit examples with fixed lepton indices below, we work in a basis where $Y_e$ is diagonal.
A simple example is $\bar u\bar d\bar d+LL\bar e$, for which
\begin{align}
    \langle\bar u_{1,1}\rangle
    =\langle\bar d_{1,2}\rangle
    =\langle\bar d_{2,3}\rangle&=a,\notag\\
    \langle L_1^\uparrow\rangle
    =\langle L_2^\downarrow\rangle
    =\langle\bar e_3\rangle&=b.
    \label{eq:udd LLe pair}
\end{align}
The first index on $\bar u$ and $\bar d$ denotes flavor and the second denotes color, and all unlisted VEVs vanish.
Both amplitudes are free at the renormalizable level.
The quark and lepton sectors are separately D-flat, while $Q=0$ and the choice of the third charged-lepton flavor make Eq.~(\ref{eq:pair F flatness}) vanish.
The canonically normalized fields are $\phi=\sqrt3a$ and $\psi=\sqrt3b$.
Their combined VEVs break all continuous MSSM gauge symmetries~\cite{Allahverdi:2006xh}.

Pairs of the same type can also satisfy the conditions.
For $Q\bar dL+Q\bar dL$, choose a unit flavor vector $q$ and two orthonormal vectors $d^{(1)},d^{(2)}$ in the null space of $q^TY_d$.
A configuration is
\begin{align}
    \langle Q_i^{1\uparrow}\rangle=aq_i,\quad
    \langle\bar d_{j1}\rangle=ad_j^{(1)},\quad
    \langle L_1^\downarrow\rangle&=a,\notag\\
    \langle Q_i^{2\downarrow}\rangle=bq_i,\quad
    \langle\bar d_{j2}\rangle=bd_j^{(2)},\quad
    \langle L_2^\uparrow\rangle&=b.
    \label{eq:QdL pair}
\end{align}
The flavor orthogonality removes the off-diagonal color D-terms, and $q^TY_dd^{(1)}=q^TY_dd^{(2)}=0$ ensures F-flatness.
The two independent lepton doublets break the weak and hypercharge generators, and the two independent color directions break the remaining color generators.

Shared fields require a joint parametrization.
For $LL\bar e+QQQL$, one may take real nonnegative amplitudes $a,b$ and
\begin{align}
    \langle L_1^\uparrow\rangle=a,\quad
    \langle L_2^\downarrow\rangle=\sqrt{a^2+b^2},\quad
    \langle\bar e_3\rangle&=a,\notag\\
    \langle Q_1^{1\uparrow}\rangle
    =\langle Q_1^{2\downarrow}\rangle
    =\langle Q_2^{3\uparrow}\rangle&=b.
    \label{eq:LLe QQQL pair}
\end{align}
This family is D-flat and F-flat for arbitrary $a,b$ and completely breaks the continuous gauge group when both are nonzero.
The shared $L_2$ component adjusts to both amplitudes.
The kinetic term induced on such a joint flat locus must be derived from the constituent fields when studying its dynamics~\cite{Enqvist:2003pb,Allahverdi:2006xh,Gumrukcuoglu:2009fj}.

Of the 36 unordered pairs formed from $\mathcal C_{\rm m}$, nine are excluded, as summarized in the first three rows of Table~\ref{tab:excluded flat pairs}.
The pair $(LL\bar e,LL\bar e)$ has no colored VEV.
Combining $LL\bar e$ with either $Q\bar dL$ or $Q\bar uL\bar e$ leaves an $SU(2)_c$ subgroup, since the colored fields of the second elementary direction occupy only one color direction.
Pairs formed entirely from $\bar u\bar d\bar d$, $\bar u\bar u\bar d\bar e$, and $\bar u\bar u\bar u\bar e\bar e$ contain no weak doublet VEV and therefore leave $SU(2)_L$ unbroken.
Moreover, with F-flatness imposed, the Higgs types give no additional allowed pair.
If $H_uH_d\ne0$, the two Higgs doublets are linearly independent.
For full-rank Yukawa matrices, $F_Q=F_{\bar u}=F_{\bar d}=0$ then requires $\bar u=\bar d=Q=0$, leaving color unbroken.
For $LH_u$ combined with a matter-field direction, $H_d=0$ and $F_Q=0$ require $\bar u=0$, excluding the five types containing $\bar u$.
In addition, $F_{\bar u}=0$ requires every nonzero $Q$ doublet to be parallel to $H_u$, which makes the weak antisymmetric contraction in $QQQL$ vanish.
The remaining combinations with $LH_u$ either leave $SU(2)_c$ unbroken or have no colored fields, as listed in Table~\ref{tab:excluded flat pairs}.

\begin{table}[t]
    \centering
    \small
    \begin{tabular}{p{0.53\linewidth}p{0.39\linewidth}}
        \toprule
        Excluded pairs & Obstruction \\
        \midrule
        $(LL\bar e,LL\bar e)$ & Unbroken $SU(3)_c$. \\[3pt]
        $(LL\bar e,Q\bar dL)$, $(LL\bar e,Q\bar uL\bar e)$ & Unbroken $SU(2)_c$. \\[3pt]
        $(\bar u\bar d\bar d,\bar u\bar d\bar d)$, $(\bar u\bar d\bar d,\bar u\bar u\bar d\bar e)$, $(\bar u\bar d\bar d,\bar u\bar u\bar u\bar e\bar e)$, $(\bar u\bar u\bar d\bar e,\bar u\bar u\bar d\bar e)$, $(\bar u\bar u\bar d\bar e,\bar u\bar u\bar u\bar e\bar e)$, $(\bar u\bar u\bar u\bar e\bar e,\bar u\bar u\bar u\bar e\bar e)$ & Unbroken $SU(2)_L$. \\[3pt]
        $(H_uH_d,X)$ for $X\in\mathcal C$ & F-flatness requires $Q=\bar u=\bar d=0$ when $H_uH_d\ne0$. \\[3pt]
        $(LH_u,X)$ for $X=\bar u\bar d\bar d$, $Q\bar uQ\bar d$, $\bar u\bar u\bar d\bar e$, $Q\bar uL\bar e$, $\bar u\bar u\bar u\bar e\bar e$ & $F_Q=0$ requires $\bar u=0$. \\[3pt]
        $(LH_u,QQQL)$ & $F_{\bar u}=0$ aligns all $Q$ doublets with $H_u$, so $QQQL=0$. \\[3pt]
        $(LH_u,Q\bar dL)$ & Unbroken $SU(2)_c$. \\[3pt]
        $(LH_u,LL\bar e)$, $(LH_u,LH_u)$ & Unbroken $SU(3)_c$. \\
        \bottomrule
    \end{tabular}
    \caption{The 28 pairs among the ten types in $\mathcal C$ excluded by requiring simultaneous F- and D-flatness together with complete gauge symmetry breaking.
}
    \label{tab:excluded flat pairs}
\end{table}

The classification above establishes the existence of simultaneous F- and D-flat configurations at the renormalizable level, but does not guarantee a stable inflationary minimum with both amplitudes nonzero.
Higher-dimensional interactions and supersymmetry-breaking terms must therefore be included when assessing whether both VEVs remain large enough to make all gauge bosons heavier than the Hubble scale.
When allowed by the gauge and discrete symmetries, nonrenormalizable superpotential operators can generate mixed contributions to the scalar potential, schematically of the form $c|a|^2|b|^8/M^6$, where $a$ and $b$ are the two amplitudes, $M$ is a cutoff scale, and $c$ is dimensionless.
Higher-dimensional K\"ahler couplings can also generate Hubble-induced mixed terms such as $c_{ab}H^2|a|^2|b|^2/M_{\mathrm{Pl}}^2$, with dimensionless $c_{ab}$.
Such interactions can stabilize one amplitude near the origin when the other develops a large VEV.
Simultaneous large VEVs therefore require an analysis of the full potential, including the stability of the stationary point and the smallest gauge boson mass.
A detailed analysis of this stabilization is beyond the scope of this work.

\section{Affleck-Dine baryogenesis and the FL argument\label{sec:ADBG}}
\label{sec:AD baryogenesis}

\subsection{Evolution and charge generation of the Affleck-Dine field}
\label{sec:AD dynamics}

We now describe how a condensate that breaks gauge symmetries during inflation can subsequently generate a baryon or lepton asymmetry.
We assume that inflation is driven by an inflaton F-term and that the AD field remains subdominant in the total energy density for simplicity.
We restrict the analysis to one flat direction described by a single canonically normalized complex field.

We adopt the standard $\mathbb Z_4^R$ symmetry, which suppresses the dangerous dimension-five proton-decay operators~\cite{Lee:2010gv,Lee:2011dya}.
As listed in Table~\ref{tab:z4r_charge}, the matter superfields have R-charge one, the Higgs superfields have R-charge zero, the superpotential has R-charge two, and the K\"ahler potential $K$ is neutral modulo four.
\begin{table}[t]
    \centering
    \begin{tabular}{c|cccccccccc}
        \hline
        Field
        & $Q$ & $\bar u$ & $\bar d$ & $L$ & $\bar e$
        & $H_u$ & $H_d$ & $\theta$ & $W$ & $K$ \\
        \hline
        $\mathbb Z_4^R$ charge
        & $1$ & $1$ & $1$ & $1$ & $1$
        & $0$ & $0$ & $1$ & $2$ & $0$ \\
        \hline
    \end{tabular}
    \caption{The $\mathbb Z_4^R$ charge assignment adopted in this work.
    Charges are understood modulo~4.
    Here, $\theta$ denotes the Grassmann superspace coordinate.}
    \label{tab:z4r_charge}
\end{table}
This forbids superpotential operators containing four matter fields, including those with one additional Higgs field, while allowing pure-matter operators of degree six, ten, and fourteen.%
\footnote{
If all generic non-renormalizable superpotential terms are allowed in the absence of the $\mathbb{Z}_4^R$ symmetry, the following discussion may no longer apply, and a sufficient baryon asymmetry may not be generated. In particular, the superpotential term that lifts and stabilizes the flat direction can arise at a lower order than the $B-L$-violating term responsible for the $A$-term. This is because an operator linear in a field with a vanishing VEV can lift the flat direction through its $F$-term, while the corresponding $A$-term vanishes along that configuration~\cite{Dine:1995kz}.
}

As an example, we consider a flavor-aligned $\bar d\bar d\bar dLL$ flat direction, along which the $F$-terms induced by superpotential operators of degree lower than ten vanish. 
The leading superpotential operator along this direction is therefore the degree-ten operator $(\bar d\bar d\bar dLL)^2$. 
We assume aligned flavor coefficients and supersymmetry-breaking terms such that this direction constitutes a stable trajectory, with all physical orthogonal modes heavier than $H$ during inflation and their subsequent excitation negligible. 
Thus, the dynamics of fields orthogonal to the flat direction can be consistently neglected. 
The degree-ten operator violates $B-L$ and can provide both radial stabilization and the charge-generating $A$-term.

We describe motion along this direction by a canonically normalized complex field $\phi$.
Writing the nonzero scalar components as $\phi_i=c_i\phi$, the normalization is $\sum_i|c_i|^2=1$.
For the configuration in Eq.~(\ref{eq:dddLL configuration}), the five nonzero coefficients are $c_i=1/\sqrt{5}$ and $\langle\phi\rangle=v$.
As a simple effective description of the lifting and charge violation, we take
\begin{align}
    W_n=\frac{\lambda\phi^n}{nM^{n-3}},
    \label{eq:Wn}
\end{align}
where $M$ is a dimensionful cutoff, distinct from the dimensionless BH mass in Sec.~\ref{subsec:RN-dS BH}, $\lambda$ is an effective dimensionless coupling, and $n$ is the degree of the operator responsible for both lifting and charge violation in this single-field description.
We retain general $n$ in the analytic expressions and use $n=10$ for the $\bar d\bar d\bar dLL$ branch in the numerical examples.

Including supersymmetry breaking, the effective potential is~\cite{Dine:1995kz}
\begin{align}
    V(\phi)&=\left(m_\phi^2-c_HH^2\right)|\phi|^2
    +\left[A(H)W_n+{\rm h.c.}\right]
    +\frac{|\lambda|^2}{M^{2n-6}}|\phi|^{2n-2},
    \label{eq:AD field full-potential}\\
    A(H)&=a_mm_{3/2}+a_HH.
\end{align}
Here, $m_\phi$ is the soft scalar mass, $m_{3/2}$ is the gravitino mass, and $a_m$ and $a_H$ are dimensionless complex coefficients.
We consider $m_{3/2}\sim m_\phi$ and $|a_m|$ of order unity, as in gravity-mediated supersymmetry breaking.
The coefficient $c_H>0$ describes a negative Hubble-induced mass squared and depends on the couplings to the inflationary sector~\cite{Dine:1995uk,Kasuya:2006wf}.
We treat it as a positive constant of order unity during inflation and the subsequent inflaton oscillation era.
For the estimates below, we assume that additional inflaton-dependent and thermal contributions are negligible, and that the field amplitude remains below the cutoff.

During inflation, $c_HH_{\rm inf}^2\gg m_\phi^2$ makes the origin unstable.
When $|A(H)|\ll H$, balancing the negative mass term against the last term in Eq.~(\ref{eq:AD field full-potential}) gives
\begin{align}
    |\phi|_{\rm min}(H)\simeq
    \left(\frac{\sqrt{c_H}}{\sqrt{n-1}|\lambda|}HM^{n-3}\right)^{1/(n-2)}\qquad {\rm for} \quad 
    c_HH^2\gg m_\phi^2.
    \label{eq: minimum AD}
\end{align}
We denote the amplitude by $v_{\rm inf}\equiv|\phi|_{\rm min}(H_{\rm inf})$.
This is the amplitude that sets the gauge boson masses during inflation, with coefficients determined by the chosen direction.

After inflation, we assume an era dominated by inflaton oscillations in a quadratic potential, so that $H\simeq2/(3t)$ until reheating.
The negative Hubble-induced term can persist during this era, and the radial amplitude decreases approximately with the moving minimum~\cite{Dine:1995kz}.
The soft mass becomes important at $H=H_{\rm osc}\sim m_\phi$, when the AD field starts to oscillate about the origin.
We assume that reheating completes after this onset.
If thermal effects do not trigger earlier oscillations~\cite{Anisimov:2000wx,Allahverdi:2000zd,Fujii:2001zr}, its amplitude at that time roughly scales as
\begin{align}
    |\phi|_{\rm osc}\sim
    \left(\frac{H_{\rm osc}}{H_{\rm inf}}\right)^{1/(n-2)}v_{\rm inf}.
    \label{eq: relatoin vev with osc}
\end{align}

To describe charge generation, write $\phi=|\phi|e^{i\theta}$ and $\delta_A(H)=\arg[A(H)\lambda]$.
The phase-dependent potential and its derivative are
\begin{align}
    V(\phi)&\supset\frac{2|A(H)\lambda|}{nM^{n-3}}|\phi|^n\cos\left(n\theta+\delta_A(H)\right),
    \label{eq: theta-potential}\\
    \frac{\partial V}{\partial\theta}
    &=-\frac{2|A(H)\lambda|}{M^{n-3}}|\phi|^n\sin\left(n\theta+\delta_A(H)\right).
    \label{eq: theta-potential derivative}
\end{align}
We focus on the case in which the Hubble-induced $A$ term is sufficiently suppressed that the phase is light during inflation~\cite{Kasuya:2008xp}.
The phase then need not coincide with the minimum selected by the soft $A$ term.
As $H$ decreases, that term becomes dynamically important and exerts a torque in the complex plane if the initial phase is misaligned.

Let $\beta_X$ be the effective charge of the canonically normalized field under a global charge $X$, such as $B$ or $L$.
For the AD field of Eq.~(\ref{eq:dddLL configuration}), $\beta_B=-1/5$ and $\beta_{B-L}=-3/5$.
The rotation generates the density
\begin{align}
    n_X=i\beta_X\left(\dot\phi^*\phi-\phi^*\dot\phi\right)
    =2\beta_X|\phi|^2\dot\theta,
\end{align}
which obeys
\begin{align}
    \dot n_X+3Hn_X&=-\beta_X\frac{\partial V}{\partial\theta},\\
    \frac{d}{d t}\left(a^3n_X\right)
    &=\frac{2\beta_Xa^3|\phi|^n}{M^{n-3}}
    \operatorname{Im}\left[A(H)\lambda e^{in\theta}\right].
    \label{eq:AD charge source}
\end{align}
The main contribution is generated near the onset of oscillation, when the torque via the soft A-term becomes effective.
Afterwards, the amplitude falls and the charge-violating term becomes negligible, leaving an approximately conserved comoving charge~\cite{Dine:1995kz,Harigaya:2014tla}.

For abundance estimates, we express this final charge as an effective density at the onset of oscillation,
\begin{align}
    n_X^{\rm osc}\equiv
    \left.\frac{a^3(t)n_X(t)}{a^3(t_{\rm osc})}\right|_{\text{after charge generation}}
    =\epsilon_Xm_\phi|\phi|_{\rm osc}^2.
    \label{eq:AD charge yield}
\end{align}
The dimensionless efficiency $\epsilon_X$ includes the charge, the initial phase, and the dynamics during the transition.
When the soft term dominates the torque, its parametric dependence is
\begin{align}
    \epsilon_X\sim\beta_X|a_m|\frac{m_{3/2}}{m_\phi}\sin\Theta_{\rm osc},\qquad
    \Theta_{\rm osc}=n\theta_{\rm osc}+\arg(a_m\lambda),
\end{align}
up to a coefficient that depends on $n$, $c_H$, and the field evolution.

As the condensate redshifts and decays, its contribution to gauge symmetry breaking disappears and its charge is transferred to ordinary particles.
Relating the generated charge to the final baryon number requires specifying whether this transfer occurs while sphaleron processes are effective~\cite{Harvey:1990qw}.
We include any such conversion in the baryon efficiency $\epsilon_B$, and assume that the condensate remains subdominant and produces no additional entropy.
The abundance estimate below uses the effective density $n_B^{\rm osc}$ defined by Eq.~(\ref{eq:AD charge yield}) after including this conversion.

\subsection{Consistency of Affleck-Dine baryogenesis with the FL argument}
\label{sec:AD consistency}

We now show how the large condensate used in AD baryogenesis can also satisfy the Higgs condition for the FL bound during inflation.
We use the single-field description and the cosmological history specified in Sec.~\ref{sec:AD dynamics}.
The gauge boson masses first constrain the inflationary amplitude $v_{\rm inf}$.
The baryon abundance then determines the required reheating temperature, while the isocurvature constraint favors a large amplitude when the phase is light during inflation.

\subsubsection{Gauge boson masses and the AD potential}

The gauge boson masses follow from Eq.~(\ref{eq:AD gauge mass matrix}) and depend on the orientation of the condensate as well as its amplitude.
For example, Eq.~(\ref{eq:dddLL configuration}) gives
\begin{align}
    m_{V,3}=\frac{g_3}{\sqrt{5}}|v|,\qquad
    m_{V,2}=\frac{g_2}{\sqrt{5}}|v|,\qquad
    m_{V,Y}=\frac{g_Y}{\sqrt{3}}|v|,
\end{align}
for the eight color, three weak, and one hypercharge gauge bosons, respectively.
Here, the non-Abelian generators satisfy $\operatorname{Tr}(T^AT^B)=\delta^{AB}/2$, and $Y_{\bar d}=1/3$ and $Y_L=-1/2$.

For a fixed direction, define the dimensionless coefficient $g_{\rm eff}$ by $m_{V,\min}=g_{\rm eff}v_{\rm inf}$, where $m_{V,\min}$ is the smallest gauge boson mass.
All couplings in this relation are evaluated during inflation.
The Higgs condition in Eq.~(\ref{eq: FLB for gauge mass}) requires
\begin{align}
    g_{\rm eff}v_{\rm inf}\gtrsim H_{\rm inf}.
    \label{eq:constraint for vev}
\end{align}
The coefficient includes the group factors and the normalization of the AD field.

Substituting Eq.~(\ref{eq: minimum AD}) gives the condition on the effective potential
\begin{align}
    \frac{c_H}{|\lambda|^2}\gtrsim
    \frac{n-1}{g_{\rm eff}^{2(n-2)}}
    \left(\frac{H_{\rm inf}}{M}\right)^{2(n-3)}.
    \label{eq: potential constraint}
\end{align}
We take the cutoff to be the reduced Planck mass, $M=M_{\mathrm{Pl}}$, in the numerical examples below.
To express this condition using the tensor-to-scalar ratio, we assume vacuum tensor fluctuations in Einstein gravity.
At the pivot scale $k_*=0.05\,\mathrm{Mpc}^{-1}$, their amplitude gives
\begin{align}
    \frac{H_*^2}{M_{\mathrm{Pl}}^2}=\frac{\pi^2}{2}A_s r
    \simeq1.04\times10^{-8}r,
    \label{eq:AD H tensor relation}
\end{align}
where $H_*$ is the Hubble parameter when the pivot scale exits the horizon and $A_s\simeq2.1\times10^{-9}$ is the scalar power spectrum~\cite{Planck:2018vyg}.
For slowly varying $H$, we identify $H_*\simeq H_{\rm inf}$.
For $M=M_{\mathrm{Pl}}$ and $n=10$, we obtain
\begin{align}
    \frac{c_H}{|\lambda|^2}\gtrsim
    7.6\times10^{-65}\left(\frac{r}{0.01}\right)^7
    \left(\frac{0.5}{g_{\rm eff}}\right)^{16}.
\end{align}
For $g_{\rm eff}=0.5$ and $c_H/|\lambda|^2$ of order unity, this condition is automatically satisfied over the range $r<0.036$ allowed by the BK18 analysis at 95\% confidence~\cite{BICEP:2021xfz}.
The Higgs condition therefore imposes no relevant additional restriction on $r$ in this regime.

Strictly speaking, 
the Higgs condition in Eq.~(\ref{eq:constraint for vev}) does not require F-flatness.
Along a D-flat direction lifted by renormalizable F-terms, the resulting quartic potential has the same radial scaling as the $n=3$ case of Eq.~(\ref{eq:AD field full-potential}).
A sufficiently small Yukawa-induced quartic coupling can therefore allow $v_{\rm inf}\gtrsim H_{\rm inf}/g_{\rm eff}$.
The standard AD baryogenesis analysis adopted here does not directly apply to such directions, so we restrict our discussion to directions that are F- and D-flat at the renormalizable level.%
\footnote{We follow Ref.~\cite{Montero:2021otb} in using $m_{\rm gauge}\gtrsim H_{\rm inf}$ as a parametric criterion for evading the FL argument through Higgsing.
Balancing a negative Hubble-induced mass squared against a D-term quartic can yield $m_{\rm gauge}\sim H_{\rm inf}$ even without D-flatness.
D-flatness is therefore not necessary to reach this mass scale, although the precise threshold for evading the FL argument requires further analysis.
Here, we focus on configurations with $m_{V,\min}\gg H_{\rm inf}$, for which all gauge fields have screening lengths well below the Hubble radius, assuming that the Higgs phase is maintained.}

\subsubsection{Baryon abundance and reheating}

We next take into account the observed baryon abundance.
During the inflaton oscillation era, both the conserved charge density and the inflaton energy density dilute as $a^{-3}$.
Using the effective density defined in Eq.~(\ref{eq:AD charge yield}), the baryon-to-entropy ratio after reheating is
\begin{align}
    Y_B\equiv\frac{n_B}{s}
    &\simeq\frac{3T_{\rm RH}}{4}
    \frac{n_B^{\rm osc}}{3M_{\mathrm{Pl}}^2H_{\rm osc}^2}
    =\frac{\epsilon_B T_{\rm RH}m_\phi|\phi|_{\rm osc}^2}
    {4M_{\mathrm{Pl}}^2H_{\rm osc}^2}.
    \label{eq:AD baryon abundance}
\end{align}
We take $\epsilon_B>0$ to obtain the observed sign of the baryon asymmetry.
It includes charge conversion as described in Sec.~\ref{sec:AD dynamics}.
For $H_{\rm osc}\sim m_\phi$, Eq.~(\ref{eq:AD baryon abundance}) reduces to $Y_B\sim\epsilon_BT_{\rm RH}|\phi|_{\rm osc}^2/(4M_{\mathrm{Pl}}^2m_\phi)$.
This estimate assumes that the condensate remains subdominant and that no additional entropy is produced after reheating for simplicity~\cite{Dine:1995kz,Harigaya:2014tla}.

Taking the observed baryon asymmetry $Y_B^{\rm obs}\simeq8.7\times10^{-11}$~\cite{Planck:2018vyg} and using Eq.~(\ref{eq: relatoin vev with osc}), the reheating temperature required for a given $v_{\rm inf}$ is
\begin{align}
    T_{\rm RH}\simeq
    \frac{4Y_B^{\rm obs}M_{\mathrm{Pl}}^2H_{\rm osc}^2}
    {\epsilon_Bm_\phi v_{\rm inf}^2}
    \left(\frac{H_{\rm inf}}{H_{\rm osc}}\right)^{2/(n-2)}.
    \label{eq:AD required reheating}
\end{align}
The reheating temperature is thus fixed by the AD field potential, the oscillation scale, and the asymmetry generation efficiency.
Substituting Eq.~(\ref{eq: minimum AD}) removes the explicit dependence on $H_{\rm inf}$.
Combining this result with Eq.~(\ref{eq:constraint for vev}) gives
\begin{align}
    T_{\rm RH}\lesssim
    \frac{4Y_B^{\rm obs}M_{\mathrm{Pl}}^2H_{\rm osc}^2g_{\rm eff}^2}
    {\epsilon_Bm_\phi H_{\rm inf}^2}
    \left(\frac{H_{\rm inf}}{H_{\rm osc}}\right)^{2/(n-2)}.
    \label{eq: RH constraint}
\end{align}
For example, taking $n=10$ and $H_{\rm osc}=m_\phi$ leads to
\begin{align}
    T_{\rm RH}\lesssim1.6\times10^7\GeV
    \left(\frac{g_{\rm eff}}{0.5}\right)^2
    \left(\frac{0.1}{\epsilon_B}\right)
    \left(\frac{m_\phi}{1\TeV}\right)^{3/4}
    \left(\frac{10^{13}\GeV}{H_{\rm inf}}\right)^{7/4},
\end{align}
where we use $M_{\mathrm{Pl}} \simeq 2.435\times10^{18}\GeV$.
At fixed $g_{\rm eff}$, $\epsilon_B$, and $m_\phi$, this upper limit scales as $r^{-7/8}$ through Eq.~(\ref{eq:AD H tensor relation}).

A larger condensate requires a lower reheating temperature to reproduce the same abundance.
Thus, when $v_{\rm inf}\gg H_{\rm inf}/g_{\rm eff}$, the required temperature lies well below the upper limit from the Higgs condition.
This condition is easily accommodated by the large amplitudes used in AD baryogenesis.
For $n=10$, $c_H=|\lambda|=1$, $m_\phi=1\TeV$, and $\epsilon_B=0.1$, Eq.~(\ref{eq:AD required reheating}) gives $T_{\rm RH}\simeq32\MeV$.%
\footnote{
For large $n$, the baryon asymmetry tends to be too large unless the reheating temperature is low. Alternatively, a modified AD mechanism, in which the AD field starts oscillating at the end of inflation, can reproduce the observed asymmetry at a higher reheating temperature without overproducing gravitinos in suitable inflation models~\cite{Yamada:2015xyr}.
}
This lies above the lower limit of a few MeV from nucleosynthesis and neutrino thermalization~\cite{deSalas:2015glj}.
We implicitly assumed charge release from the condensate without excessive dilution or washout.
If charge is released after sphaleron freeze-out, the $\bar d\bar d\bar dLL$ condensate can instead supply its directly generated baryon charge.

\subsubsection{Isocurvature constraint}

When the phase of the AD field is light during inflation, its quantum fluctuations can generate baryon isocurvature perturbations~\cite{Enqvist:1998pf,Enqvist:1999hv,Kasuya:2008xp}.
We write the canonical complex AD field as $\phi=\rho e^{i\theta}/\sqrt{2}$. 
The real radial field has $\rho_* =\sqrt{2}v_*$, where $v_*=|\phi_*|$ at horizon exit.
The power spectrum of the phase fluctuation is given by 
\begin{align}
    \mathcal P_{\delta\theta}^{1/2}(k_*)
    =\frac{H_*}{2\pi\rho_*}
    =\frac{H_*}{2\pi\sqrt{2}v_*}.
\end{align}
The baryon isocurvature depends on the phase sensitivity of $\epsilon_B$. 

If phase evolution before the onset of oscillation is negligible, $\theta_{\rm osc}\simeq\theta_*$ and $\Theta_{\rm osc}\simeq\Theta\equiv n\theta_*+\arg(a_m\lambda)$.
We then expect $\epsilon_B\propto\sin\Theta$ and obtain 
\begin{align}
    S_{b\gamma}\equiv\delta\ln Y_B\simeq n\cot\Theta\,\delta\theta,
\end{align}
and its power spectrum is 
\begin{align}
    \mathcal P_{S_{b\gamma}}(k_*)
    =\left(\frac{n\cot\Theta\,H_*}{2\pi\sqrt{2}v_*}\right)^2.
    \label{eq:AD baryon isocurvature}
\end{align}
More generally, $n\cot\Theta$ is replaced by $\partial\ln\epsilon_B/\partial\theta_*$, including any phase evolution before charge generation~\cite{Kasuya:2008xp,Harigaya:2014tla}.

Identifying $v_*\simeq v_{\rm inf}$, the Higgs condition gives
\begin{align}
    \mathcal P_{S_{b\gamma}}^{1/2}(k_*)
    \lesssim\frac{n|\cot\Theta|g_{\rm eff}}{2\pi\sqrt{2}}
    \simeq0.56\left(\frac{n}{10}\right)|\cot\Theta|
    \left(\frac{g_{\rm eff}}{0.5}\right).
\end{align}
Remarkably, this condition is independent of both the VEV of the AD field and the inflationary scale.
For a nearly scale-invariant baryon isocurvature mode uncorrelated with the adiabatic perturbation, with no intrinsic dark matter isocurvature, we use the approximate observational limit
\begin{align}
    \mathcal P_{S_{b\gamma}}^{1/2}(k_*)\lesssim S_{\max},\qquad
    S_{\max}=5\times10^{-5}.
\end{align}
This value includes the conversion between baryon isocurvature and the equivalent cold dark matter mode constrained by CMB data~\cite{Harigaya:2014tla,Planck:2018jri}.
The resulting lower limit on the condensate is
\begin{align}
    v_{\rm inf}\gtrsim
    \frac{n|\cot\Theta|H_{\rm inf}}{2\pi\sqrt{2}S_{\max}}.
    \label{eq:AD isocurvature amplitude bound}
\end{align}
For $n=10$, this requires $v_{\rm inf}/H_{\rm inf}\gtrsim2.3\times10^4|\cot\Theta|$.
For $|\cot\Theta|$ of order unity, this amplitude is far above the Higgs threshold $v_{\rm inf}/H_{\rm inf}\gtrsim2$ for $g_{\rm eff}=0.5$.
The Higgs condition is therefore automatically satisfied once the observational isocurvature limit is imposed in this regime.

For the single-direction monomial types listed in Table~\ref{tab: Flat direction}, the operators $I^2$ correspond to $n=10$ or $14$.
Taking $M=M_{\mathrm{Pl}}$ and $H_{\rm inf}\simeq H_*$, Eq.~(\ref{eq: minimum AD}) gives
\begin{align}
    \frac{v_{\rm inf}}{H_*}
    &\simeq 4.5\times10^4
    \left(\frac{c_H}{|\lambda|^2}\right)^{1/16}
    \left(\frac{10^{13}\,\mathrm{GeV}}{H_*}\right)^{7/8}
    \qquad {\rm for} \quad n=10,
    \nonumber\\
    \frac{v_{\rm inf}}{H_*}
    &\simeq 7.8\times10^4
    \left(\frac{c_H}{|\lambda|^2}\right)^{1/24}
    \left(\frac{10^{13}\,\mathrm{GeV}}{H_*}\right)^{11/12}
    \qquad  {\rm for} \quad n=14.
\end{align}
For comparison, the commonly studied cases with $n=4$ and $6$ give
\begin{align}
    \frac{v_{\rm inf}}{H_*}
    &\simeq 3.7\times10^2
    \left(\frac{c_H}{|\lambda|^2}\right)^{1/4}
    \left(\frac{10^{13}\,\mathrm{GeV}}{H_*}\right)^{1/2}
    \qquad  {\rm for} \quad n=4,
    \nonumber\\
    \frac{v_{\rm inf}}{H_*}
    &\simeq 9.0\times10^3
    \left(\frac{c_H}{|\lambda|^2}\right)^{1/8}
    \left(\frac{10^{13}\,\mathrm{GeV}}{H_*}\right)^{3/4}
    \qquad  {\rm for} \quad  n=6.
\end{align}
The larger powers therefore allow a larger condensate and suppress the phase fluctuations more efficiently~\cite{Harigaya:2014tla}.
For $H_*=10^{13}\,\mathrm{GeV}$ and $c_H=|\lambda|=1$, Eq.~(\ref{eq:AD baryon isocurvature}) yields
\begin{align}
    \mathcal P_{S_{b\gamma}}^{1/2}
    &\simeq 2.5\times10^{-5}|\cot\Theta|,
    \qquad n=10,
    \nonumber\\
    \mathcal P_{S_{b\gamma}}^{1/2}
    &\simeq 2.0\times10^{-5}|\cot\Theta|,
    \qquad n=14.
\end{align}
Both cases satisfy the adopted isocurvature limit for $|\cot\Theta|\simeq1$, without requiring a small effective coupling $\lambda$.
By comparison, the same parameters give $1.2\times10^{-3}|\cot\Theta|$ and $7.5\times10^{-5}|\cot\Theta|$ for $n=4$ and $6$, respectively.
Thus, for the parameters considered here, higher-order lifting operators help satisfy the baryon isocurvature constraint, as illustrated by the $n=10$ branch assumed in Sec.~\ref{sec:AD dynamics}.

\subsubsection{Summary of constraints}

We summarize the formal upper bounds implied by the Higgs condition in Table~\ref{tab:constraints_n}.
The $n=10$ example follows from the $\bar d\bar d\bar dLL$ condensate, while 
the other rows illustrate the dependence of the general effective expressions on $n$ without asserting a corresponding MSSM realization.
The bound on $T_{\rm RH}$ becomes more stringent as $n$ increases and formally approaches $52\TeV$.

\begin{table}[htbp]
\centering
\renewcommand{\arraystretch}{1.3}
\begin{tabular}{p{0.1\linewidth}p{0.25\linewidth}p{0.25\linewidth}p{0.16\linewidth}}
\toprule
$n$
& $r$
& $T_{\rm RH}$
& $\mathcal{P}_{S_{b\gamma}}^{1/2}$ \\
\midrule
6  & $\lesssim 8.9\times10^{6}$ & $\lesssim 5.2\times10^{9}\,{\rm GeV}$ & $\lesssim 0.34\,|\cot\Theta|$ \\
7  & $\lesssim 1.1\times10^{7}$ & $\lesssim 5.2\times10^{8}\,{\rm GeV}$ & $\lesssim 0.39\,|\cot\Theta|$ \\
8  & $\lesssim 1.2\times10^{7}$ & $\lesssim 1.1\times10^{8}\,{\rm GeV}$ & $\lesssim 0.45\,|\cot\Theta|$ \\
9  & $\lesssim 1.4\times10^{7}$ & $\lesssim 3.7\times10^{7}\,{\rm GeV}$ & $\lesssim 0.51\,|\cot\Theta|$ \\
10 & $\lesssim 1.4\times10^{7}$ & $\lesssim 1.6\times10^{7}\,{\rm GeV}$ & $\lesssim 0.56\,|\cot\Theta|$ \\
$\rightarrow\infty$ & $\rightarrow2.4\times 10^7$ & $\rightarrow52\TeV$ & $\rightarrow\infty$ \\
\bottomrule
\end{tabular}
\caption{Formal upper bounds from the Higgs condition for different values of $n$.
We take $M=M_{\mathrm{Pl}}$, $c_H=|\lambda|=1$, and $g_{\rm eff}=0.5$, with $H_{\rm inf}=10^{13}\GeV$, $H_{\rm osc}=m_\phi=1\TeV$, and $\epsilon_B=0.1$ for the reheating bounds.
The $r$ bounds are formal extrapolations and impose no restriction in the physical regime $H_{\rm inf}\ll M_{\mathrm{Pl}}$.
The $n\to\infty$ row is only an algebraic limit of the effective expressions.}
\label{tab:constraints_n}
\end{table}

\begin{figure}
    \centering
    \includegraphics[width=\linewidth]{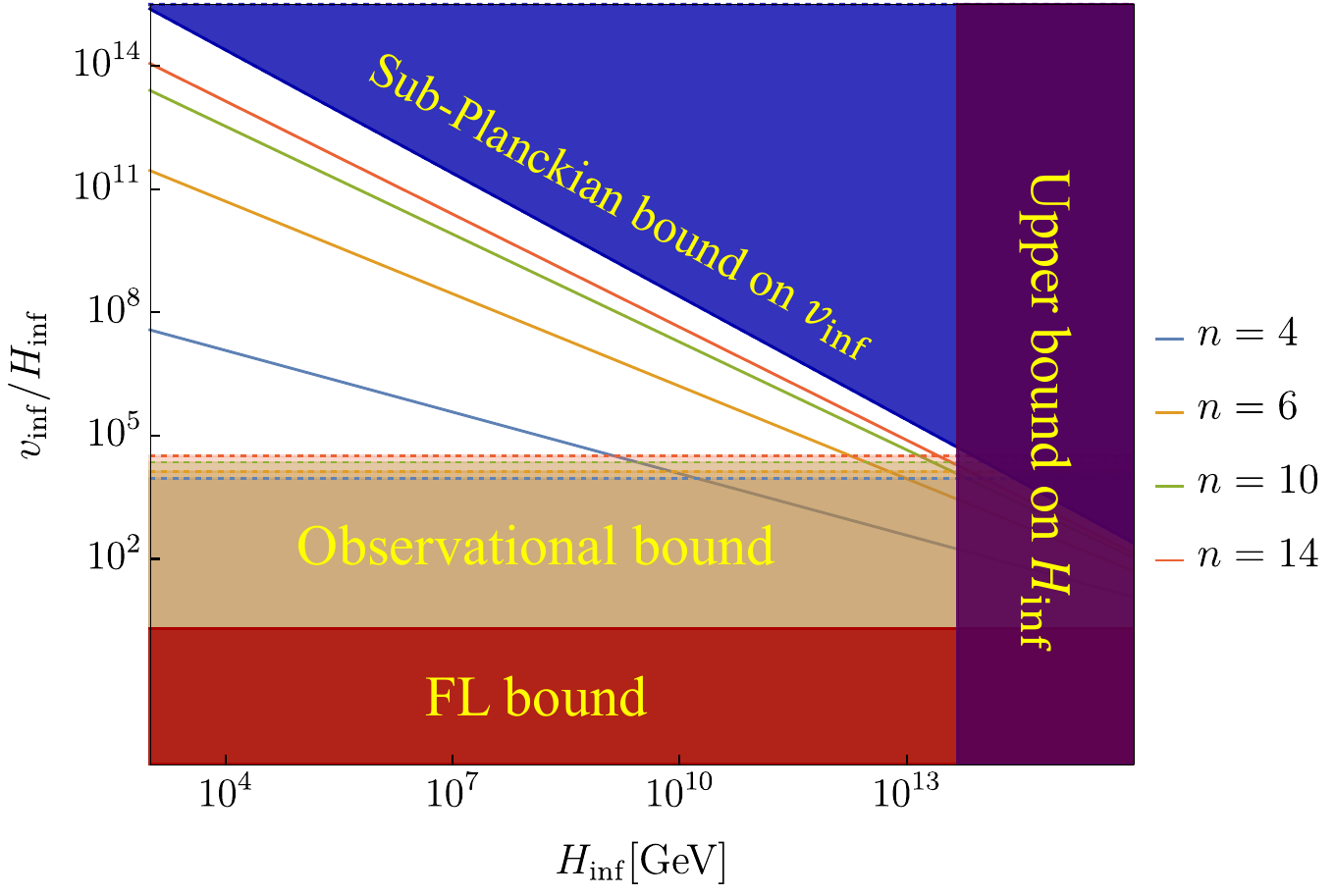}
    \caption{The ratio $v_{\rm inf}/H_{\rm inf}$ as a function of $H_{\rm inf}\simeq H_*$ from Eq.~(\ref{eq: minimum AD}), for $M=M_{\mathrm{Pl}}$ and $c_H=|\lambda|=1$.
    The solid curves correspond to the indicated values of $n$.
    The red shaded region violates the Higgs condition in Eq.~(\ref{eq:constraint for vev}) for $g_{\rm eff}=0.5$.
    The horizontal dashed lines mark the isocurvature lower bounds in Eq.~(\ref{eq:AD isocurvature amplitude bound}) for the corresponding values of $n$, with $|\cot\Theta|=1$ and $S_{\max}=5\times10^{-5}$.
    The region below each line is excluded under the light-phase assumption.
    The purple shaded region is excluded by the BK18 bound $r<0.036$, corresponding to $H_{\rm inf}\lesssim4.7\times10^{13}\,\mathrm{GeV}$ through Eq.~(\ref{eq:AD H tensor relation}).
    The blue shaded region violates the requirement $v_{\rm inf}<M_{\mathrm{Pl}}$.
    }
    \label{fig:vInfOverH}
\end{figure}

Figure~\ref{fig:vInfOverH} shows $v_{\rm inf}/H_{\rm inf}$ as a function of $H_{\rm inf}$, as obtained from Eq.~(\ref{eq: minimum AD}).
We compare the Higgs condition in Eq.~(\ref{eq:constraint for vev}) with the baryon isocurvature bound in Eq.~(\ref{eq:AD isocurvature amplitude bound}), the observational upper limit on $H_{\rm inf}$, and the requirement $v_{\rm inf}<M_{\mathrm{Pl}}$.
As $n$ increases, the curves shift upward and their logarithmic slopes become more negative.
In the formal limit $n\to\infty$, they approach $v_{\rm inf}=M_{\mathrm{Pl}}$, where the expansion in higher-dimensional operators is no longer controlled.
For the light phase and $|\cot\Theta|=1$ assumed in the figure, the isocurvature lower bound on the amplitude is approximately four orders of magnitude stronger than the Higgs condition.
Consequently, configurations satisfying the isocurvature bound also satisfy the Higgs condition by a wide margin.
Within these assumptions, AD baryogenesis with complete gauge symmetry breaking avoids the severe FL restrictions discussed in Sec.~\ref{sec:duration} and Sec.~\ref{sec:inflation}, though the baryon abundance and thermal-history requirements must still be checked separately.

Within the assumptions of Sec.~\ref{sec:AD dynamics}, the Higgs mechanism provides a way to satisfy the FL bound that is particularly well suited to AD baryogenesis.
The large condensate can generate the observed baryon abundance and suppress isocurvature while readily making all gauge bosons heavier than the Hubble scale.
Although the resulting FL conditions on the baryon abundance, the tensor-to-scalar ratio, and isocurvature introduce no significant additional restrictions in the regime illustrated above, complete gauge symmetry breaking restricts the allowed condensate configurations.
Within the renormalizable F- and D-flat condition, only five of the 28 monomial types admit complete breaking individually, as shown in Table~\ref{tab: Flat direction}, and eight types enter the allowed pairs.
All five single-direction monomials carry nonzero $B-L$.
Although the lifting operators, charge transfer, and thermal history may affect the detailed realization~\cite{Allahverdi:2005mz,Allahverdi:2008pf,Mukaida:2012qn}, the FL requirement is naturally accommodated by the same large VEV that supports successful AD baryogenesis.

\section{Discussion and conclusions}
\label{sec:discussion}

We have examined how supersymmetric flat directions can reconcile inflation with the FL bound.
We have identified individual MSSM flat directions and combinations of two directions that can completely break the continuous gauge group during inflation.
With suitable charge-violating interactions, these condensates can also generate the baryon asymmetry through the AD mechanism.
In particular, we find that requiring a single flat direction to break all gauge symmetries naturally selects flat directions carrying a nonzero $B-L$ charge.
This compatibility provides a motivation for AD baryogenesis from the FL bound, with the same large VEV responsible for both inflationary gauge symmetry breaking and the subsequent generation of the baryon asymmetry.

Complete breaking of the gauge symmetry generally requires several field components within a single flat direction or a combination of multiple flat directions.
For a single direction, the leading nonrenormalizable operator that stabilizes the condensate may involve a high power of the field.
The $\bar d\bar d\bar dLL$ flat direction provides an example, as it can be lifted by an operator of degree ten.
The resulting suppression of the lifting potential can allow a large inflationary VEV, helping to satisfy the baryon isocurvature constraint even for high-scale inflation.

The scalar condensate may fragment into Q-balls after inflation~\cite{Kusenko:1997si,Enqvist:1997si,Kasuya:1999wu,Kasuya:2000wx}.
Their formation, evaporation, and decay can modify charge release and sphaleron conversion, and may produce additional entropy~\cite{Fujii:2002kr,Kamada:2012bk,Harigaya:2014tla}.
The baryon abundance must then be evaluated using the corresponding charge transfer and dilution history.
These later processes do not alter the inflationary gauge masses.
Thus, although the abundance calculation changes, the large inflationary VEVs considered here remain compatible with the FL requirement.

For multiple flat directions, the coupled evolution can modify the timing and efficiency of charge generation~\cite{Senami:2002kn,Enqvist:2003pb,Kamada:2008sv}, as well as the resulting isocurvature perturbations~\cite{Kamada:2008sv}.
A dedicated analysis of charge generation and perturbations in the configurations that completely break the gauge symmetries is left for future work.
Such an analysis would extend the connection between the FL bound and AD baryogenesis beyond the single-field limit.
The distinctive cosmological and phenomenological consequences of the flat directions identified here also warrant further study.

\section*{Acknowledgements}
This work was supported by JSPS KAKENHI Grant Number 23K13092.
This work was supported by Graduate Program on Physics for the Universe (GP-PU), Tohoku University (KN).


\bibliography{reference}

\end{document}